\documentclass[superscriptaddress,aps,preprintnumbers,amsmath,showpacs,amssymb,prd,nofootinbib,preprint]{revtex4-1}

\usepackage[dvipdfmx]{graphicx}
\usepackage{amsfonts}
\usepackage[mathscr]{eucal}
\usepackage{dcolumn}
\usepackage{bm}
\usepackage[colorlinks=true,linkcolor=blue,citecolor=blue]{hyperref}
\usepackage{hyperref}
\usepackage{color}

\begin{document}


\newcommand{\vev}[1]{ \left\langle {#1} \right\rangle }
\newcommand{\bra}[1]{ \langle {#1} | }
\newcommand{\ket}[1]{ | {#1} \rangle }
\newcommand{\eV}{ \ {\rm eV} }
\newcommand{\KeV}{ \ {\rm keV} }
\newcommand{\MeV}{\  {\rm MeV} }
\newcommand{\GeV}{\  {\rm GeV} }
\newcommand{\TeV}{\  {\rm TeV} }
\newcommand{\1}{\mbox{1}\hspace{-0.25em}\mbox{l}}
\newcommand{\Red}[1]{{\color{red} {#1}}}

\newcommand{\lmk}{\left(}  
\newcommand{\rmk}{\right)}
\newcommand{\lkk}{\left[}  
\newcommand{\rkk}{\right]}
\newcommand{\lhk}{\left \{ }  
\newcommand{\rhk}{\right \} }
\newcommand{\del}{\partial}  
\newcommand{\la}{\left\langle} 
\newcommand{\ra}{\right\rangle}
\newcommand{\half}{\frac{1}{2}}

\newcommand{\bea}{\begin{array}}
\newcommand{\eea}{\end{array}}
\newcommand{\beq}{\begin{eqnarray}}
\newcommand{\eeq}{\end{eqnarray}}

\newcommand{\dd}{\mathrm{d}}
\newcommand{\Mpl}{M_{\rm Pl}}
\newcommand{\mg}{m_{3/2}}
\newcommand{\abs}[1]{\left\vert {#1} \right\vert}
\newcommand{\mphi}{m_{\phi}}
\newcommand{\Hz}{\ {\rm Hz}}
\newcommand{\for}{\quad \text{for }}
\newcommand{\Min}{\text{Min}}
\newcommand{\Max}{\text{Max}}
\newcommand{\Kahler}{K\"{a}hler }
\newcommand{\cphi}{\varphi}
\newcommand{\Tr}{\text{Tr}}
\newcommand{\diag}{{\rm diag}}

\newcommand{\SUf}{SU(3)_{\rm f}}
\newcommand{\Upq}{U(1)_{\rm PQ}}
\newcommand{\Zpq}{Z^{\rm PQ}_3}
\newcommand{\Cpq}{C_{\rm PQ}}
\newcommand{\ubar}{u^c}
\newcommand{\dbar}{d^c}
\newcommand{\ebar}{e^c}
\newcommand{\nubar}{\nu^c}
\newcommand{\Ndw}{N_{\rm DW}}
\newcommand{\Fpq}{F_{\rm PQ}}
\newcommand{\fpq}{v_{\rm PQ}}
\newcommand{\Br}{{\rm Br}}
\newcommand{\Lag}{\mathcal{L}}
\newcommand{\Lqcd}{\Lambda_{\rm QCD}}

\newcommand{\ji}{j_{\rm inf}} 
\newcommand{\jb}{j_{B-L}} 
\newcommand{\M}{M} 
\newcommand{\im}{{\rm Im} }
\newcommand{\re}{{\rm Re} }

\def\lrf#1#2{ \left(\frac{#1}{#2}\right)}
\def\lrfp#1#2#3{ \left(\frac{#1}{#2} \right)^{#3}}
\def\lrp#1#2{\left( #1 \right)^{#2}}
\def\REF#1{Ref.~\cite{#1}}
\def\SEC#1{Sec.~\ref{#1}}
\def\FIG#1{Fig.~\ref{#1}}
\def\EQ#1{Eq.~(\ref{#1})}
\def\EQS#1{Eqs.~(\ref{#1})}
\def\TEV#1{10^{#1}{\rm\,TeV}}
\def\GEV#1{10^{#1}{\rm\,GeV}}
\def\MEV#1{10^{#1}{\rm\,MeV}}
\def\KEV#1{10^{#1}{\rm\,keV}}
\def\blue#1{\textcolor{blue}{#1}}
\def\red#1{\textcolor{blue}{#1}}

\newcommand{\fa}{f_{a}}
\newcommand{\Uh}{U(1)$_{\rm H}$}
\newcommand{\osc}{_{\rm osc}}
\newcommand{\bear}{\begin{array}}  
\newcommand {\eear}{\end{array}}

\newcommand{\mav}{\left. m_a^2 \right\vert_{T=0}}
\newcommand{\mat}{m_{a, {\rm QCD}}^2 (T)}
\newcommand{\mam}{m_{a, {\rm M}}^2 }


\preprint{
TU-1012,\, \\
IPMU15-0212,\, \\
APCTP Pre2015-029 \\
}
\title{
The QCD Axion from  Aligned Axions  and Diphoton Excess 
}
\author{
 Tetsutaro Higaki
}
\affiliation{Department of Physics, Keio University, Kanagawa 223-8522, Japan}
\author{
Kwang Sik Jeong
}
\affiliation{Department of Physics, Pusan National University, Busan, 609-735, Korea
}
\author{
Naoya Kitajima
}
\affiliation{Asia Pacific Center for Theoretical Physics, Pohang 790-784, Korea}
\author{
Fuminobu Takahashi
}
\affiliation{Department of Physics, Tohoku University, 
Sendai, Miyagi 980-8578, Japan} 
\affiliation{Kavli IPMU (WPI), UTIAS, 
The University of Tokyo, 
Kashiwa, Chiba 277-8583, Japan}


\begin{abstract} 

We argue that the QCD axion can arise from many aligned axions with  decay constants much smaller than
the conventional axion window. If the typical decay constant is of ${\cal O}{(100)}$~GeV to $1$~TeV,
one or more of the axions or saxions may account for the recently found diphoton excess at $\sim 750$~GeV.
Our scenario predicts many axions and saxions coupled to gluons with decay constants of order the weak scale, and
therefore many collider signatures by heavy axions and saxions will show up at different energy scales.
In particular, if the inferred broad decay width is due to multiple axions or saxions,
a non-trivial peak structure may become evident when more data is collected.  
We also discuss cosmological implications of the aligned QCD axion scenario.
In the Appendix we give a possible UV completion and argue that  the high quality of the Peccei-Quinn symmetry is
naturally explained in our scenario.

\end{abstract}

\maketitle

The ATLAS and CMS experiments at the Large Hadron Collider recently announced that
they observed an excess in the diphoton resonance search at $\sim 750$\,GeV with $2-3 \sigma$
level~\cite{diphoton}. The excess may be interpreted as a new particle
decaying into two photons. Among various theoretical possibilities, a heavy axion field
is an interesting and promising candidate~\cite{Harigaya:2015ezk,Mambrini:2015wyu,Backovic:2015fnp,Angelescu:2015uiz,Nakai:2015ptz,
Knapen:2015dap,Buttazzo:2015txu,Pilaftsis:2015ycr,Franceschini:2015kwy,DiChiara:2015vdm}.\footnote{Cosmological 
and collider experimental signatures of such heavy axions were studied in Ref.~\cite{Jaeckel:2012yz}.
}
Then, the question is why such heavy axion exists in nature.
The purpose of this letter is to point out that many axions with the decay constant at the weak scale
may conspire together to form the QCD axion with an axion decay constant $f_{\rm QCD} \gtrsim 10^{9}$\,GeV by
the alignment mechanism~\cite{Kim:2004rp}. 
Such multiple axions have been studied in the so-called axiverse scenario~\cite{Arvanitaki:2009fg,Acharya:2010zx} 
and the alignment mechanism~\cite{Choi:2014rja,Higaki:2014pja}.
In particular, multiple axions naturally form axion landscape~\cite{Higaki:2014pja,Higaki:2014mwa}
if the number of shift symmetry breaking terms is greater than the number of axions.\footnote{See 
also Refs.~\cite{Wang:2015rel, Masoumi:2016eqo} for recent studies on the vacuum selection and stability 
in the axion landscape.}

The Peccei-Quinn (PQ) mechanism solves the strong CP problem 
by promoting the strong CP phase $\theta$ to a dynamical variable, the axion $a_{\rm QCD}$~\cite{Peccei:1977hh,Weinberg:1977ma} 
(see Refs.~\cite{Kim:2008hd,Bae:2008ue,Ringwald:2012hr,Kawasaki:2013ae} for recent reviews).
The conventional axion window for the axion decay constant $f_{\rm QCD}$ is given by
$\GEV{9} \lesssim f_{\rm QCD} \lesssim \GEV{12}$, where the lower bound comes from the star/supernova cooling argument,
and the upper bound from the overabundance of the axion dark matter with the initial misalignment angle of order unity.
The origin of the axion decay constant in the intermediate scale is a puzzle, and it depends on how the QCD axion
arises. In a field theoretic QCD axion model, the axion decay constant is determined by the vacuum expectation value
of the PQ scalar. Alternatively, the QCD axion may have a stringy origin,
for which the natural scale of the axion decay constant is the string scale. 

In this letter we point out a possibility that the QCD axion arises from a combination of many axions with decay
constants much smaller than the conventional axion window, based on the so-called alignment 
mechanism~\cite{Kim:2004rp}.  
As was noticed in Ref.~\cite{Choi:2014rja}, implemented by many axions, the alignment mechanism can 
exponentially enhance the effective axion decay constant without introducing extremely large charges.
The alignment with many axions was discussed further in Refs.~\cite{Higaki:2014pja,Harigaya:2014rga,Higaki:2014mwa,Choi:2015fiu,Kaplan:2015fuy,Wang:2015rel}, 
and the application of the alignment mechanism to the QCD axion was also considered
in Ref.~\cite{Higaki:2014qua}.
If this is the case, there are many axions and saxions in the low energy, some of which may be within the reach of
collider experiments such as LHC. Interestingly, the ATLAS and CMS experiments have found an excess
in the diphoton resonance search at about $750$\,GeV. The excess may be due to the decay of one or more of heavy axions
needed to form the QCD axion by the alignment mechanism. Our scenario predicts that many
other excesses in the diphoton resonance search will show up at different energy scales because there must be
at least of order $10$ such heavy axions. Alternatively, it is similarly possible that the observed diphoton excess is due to
one of the saxions, and in this case, the axions can be lighter and searched for by different techniques.

To implement the axion alignment mechanism, we consider a hidden sector with multiple periodic axions,
\beq
\phi_i \equiv \phi_i + 2\pi f_i, 
\eeq
where $i=1,2,..,N$, and the axions are assumed to have a similar decay constant
\beq
f_i \sim f.
\eeq 
Then the alignment mechanism can make one of the axions have an exponentially enhanced 
effective decay constant~\cite{Choi:2014rja,Choi:2015fiu,Kaplan:2015fuy}:
\beq
f_{\rm QCD} \sim e^{\xi N} f, 
\eeq
where $\xi={\cal O}(1)$. We will identify this axion with the QCD axion, $a_{\rm QCD}$, that solves the strong CP problem.
An enhanced effective decay constant can be achieved, for instance, in the simple model with 
the interactions~\cite{Choi:2014rja,Choi:2015fiu}\footnote{
See also Ref.~\cite{Harigaya:2014rga} where  a similar set-up was given in the linear representation.
}
\beq
\label{action}
\Delta {\cal L} = 
\sum^{N-1}_{i=1} \Lambda^4_i \cos\left(
\frac{\phi_i}{f_i} + n_i \frac{\phi_{i+1}}{f_{i+1}} \right)
+ \frac{g^2_3}{32\pi^2}\frac{k_s \phi_N}{f_N} 
G^{\mu\nu}\tilde G_{\mu\nu}
+ \frac{g^2_1}{32\pi^2}\frac{k \phi_N}{f_N} 
B^{\mu\nu}\tilde B_{\mu\nu},
\eeq
for $\Lambda_i \gg \Lambda_{\rm QCD}$, with $n_i$, $k_s$ and $k$ being integers
which parameterize the discrete degrees of freedom in the underlying nonperturbative dynamics
responsible for the axion potential. 
We give one possible UV completion in the Appendix.
Here $B^{\mu\nu}$ and $G^{\mu\nu}$ are the U$(1)_Y$ and SU$(3)_c$ field strength,
respectively, and the tilded ones are their dual tensor.
In the model we have assumed that $\phi_N$ does not couple to the SU$(2)_L$ gauge bosons, which 
would be the case when the PQ quarks are singlet under SU$(2)_L$.
As noticed in Refs.~\cite{Choi:2014rja,Choi:2015fiu}, the model has a flat direction composed as
\beq
a_{\rm QCD} \propto \sum^N_{i=1} (-1)^{i-1} 
\left(\prod^{N-1}_{j=1} n_j \right) f_i \phi_i,
\eeq
which obtains a mass from the QCD instanton effects. 
The effective action for $a_{\rm QCD}$ reads
\beq
{\cal L}_{\rm eff} 
= \frac{g^2_3}{32\pi^2}\frac{k_s a_{\rm QCD}}{f_{\rm QCD} } 
G^{\mu\nu}\tilde G_{\mu\nu}
+ \frac{g^2_1}{32\pi^2}\frac{k a_{\rm QCD}}{f_{\rm QCD} } 
B^{\mu\nu}\tilde B_{\mu\nu},
\eeq
in the canonical basis, where the effective axion constant is given by
\beq
f_{\rm QCD} = \sqrt{\sum^N_{i=1} 
\left(\prod^{N-1}_{j=1} n_j \right) 
f^2_i}\,
\sim e^{\xi N} f,
\eeq
with $\xi={\cal O}(1)$.  In order to enhance the effective decay constant by a factor of $10^6$, we need $N \simeq 10$ or more
axions.

In the scenario under consideration, there are $N-1$ axions much heavier than the QCD axion, and their decay constants
are of the order  $f$. 
Let us consider the case where one of them, $a_{\rm hid}$, has mass $m_{\rm hid}=750$~GeV and 
decay constant $f_{\rm hid} \sim f$. 
The effective couplings of $a_{\rm hid}$ can be easily read off from the action (\ref{action}): 
\beq
{\cal L}_{\rm eff} =
\frac{g^2_3}{32\pi^2}\frac{k_s a_{\rm hid}}{f_{\rm hid}} 
G^{\mu\nu}\tilde G_{\mu\nu}
+ \frac{g^2_1}{32\pi^2}\frac{k a_{\rm hid}}{f_{\rm hid}} 
B^{\mu\nu}\tilde B_{\mu\nu}
+
m^2_{\rm hid} f^2_{\rm hid} \cos\left(\frac{a_{\rm hid}}{f_{\rm hid}}\right),
\eeq
where we have omitted a mixing parameter which is considered to be of order unity.
The axion $a_{\rm hid}$ is produced via gluon fusion process, and decays into SM gauge bosons through the above interactions. 
Note that the decay rates are given
\beq
\Gamma_{a_{\rm hid}\to \gamma\gamma}:
\Gamma_{a_{\rm hid}\to Z\gamma}:
\Gamma_{a_{\rm hid}\to ZZ} 
\simeq 1:2\tan^2\theta_W:\tan^4\theta_W
=
1:0.6:0.08,
\eeq
with  
\beq
\Gamma_{a_{\rm hid}\to \gamma\gamma}
= \frac{g^4_1 \cos^4\theta_W}{4(4\pi)^5} \left(\frac{k}{f_{\rm hid}}\right)^2 m^3_{\rm hid}
\simeq 0.3{\rm MeV}
\left(\frac{m_{\rm hid}}{750{\rm GeV}}\right)^3
\left(\frac{f_{\rm hid}/k}{100{\rm GeV}}\right)^{-2}.
\eeq
So there is a mild tension with the constraint on the $Z \gamma$ channel at the $8$\,TeV LHC run \cite{Aad:2014fha}
if the excess is due to the axion $a_{\rm hid}$.
The production cross section for $a_{\rm hid}$ is estimated to be~\cite{Buttazzo:2015txu}
\beq
\sigma(pp\to a_{\rm hid} ) |_{8{\rm TeV}} &\approx& 1.5\,{\rm pb}
 \left(\frac{f_{\rm hid} /k_s}{100{\rm GeV}}\right)^{-2},
\nonumber \\
\sigma(pp\to a_{\rm hid}) |_{13{\rm TeV}} &\approx& 6.9\,{\rm pb}
\left(\frac{f_{\rm hid} /k_s}{100{\rm GeV}}\right)^{-2},
\eeq
for the 8 and 13 TeV measurements at the LHC, respectively.
Because the branching ratio into diphoton is given by
\beq
{\Br}(a_{\rm hid} \to \gamma\gamma) \simeq 
\frac{\Gamma_{a_{\rm hid} \to \gamma\gamma}}{\Gamma_{a_{\rm hid}\to gg}}
\simeq 
0.88\times 10^{-3}
\left(\frac{k}{k_s}\right)^2,
\eeq
one can find
\beq
\sigma(pp\to a_{\rm hid}) |_{13{\rm TeV}}
\times {\Br}(a_{\rm hid} \to \gamma\gamma) 
  &\approx&
6.1\,{\rm fb} 
 \left(\frac{f_{\rm hid}/k}{100{\rm GeV}}\right)^{-2}.
\eeq
Hence the hidden axion can account for the observed diphoton excess at $750$~GeV.
For instance, we may take $f_{\rm hid} \sim 1$\,TeV with $k = 10$.
It is also worth noting that the production cross section times branching ratio for the process $pp \to a_{\rm hid} \to \gamma\gamma$
is approximately independent of $k_s$ because in our scenario the axion is produced by gluon fusion and dominantly decays
into gluons.  
To explain the diphoton excess, we need $f_{\rm hid}$ around $k\times 100$~GeV, which can be above TeV for large $k$.
Such large $k$ may give information on the PQ sector as $k_s$ corresponds to the number of PQ quark pairs and $k$ is roughly 3 times larger than $k_s$
for PQ quarks carrying an electric charge of order unity.
A large $k$ may indicate that there are also PQ leptons having masses around $f_{\rm hid}$.

Note that our aligned QCD axion scenario requires many axions, some of which may have masses close to each other.
This raises a possibility that multiple axions or saxions contribute to the diphoton excess, in which case the
inferred broad width may be due to multiple peaks. If this is the case, a non-trivial peak structure may show up
when more data is collected in the rest of LHC Run-2.
Another interesting feature is that there can be dark radiation from the hidden sector.
The potential (\ref{action}) for multiple axions can be generated by hidden strong interactions.
In such scenario a plausible possibility is that the hidden sector also possesses unbroken Abelian or weakly coupled non-Abelian gauge 
groups, and then the hidden sector can give a sizable contribution to (self-interacting) dark radiation~\cite{Jeong:2013eza}.

So far we have focused on the mixings between axions and implications for the diphoton excess.
In the UV completion, there also exists a saxion $s_i$ for each axion $\phi_i$, and the saxions are
generically coupled to both gluons and axions with decay constants $f_i$. In particular, 
there is no special reason to expect the alignment to occur for the saxion mixings.
If one of the saxions, $s_{\rm hid}$,  has a coupling like
\beq
{\cal L}_{\rm eff} =
\frac{g^2_3}{32\pi^2}\frac{k_s s_{\rm hid}}{f_{\rm hid}} 
G^{\mu\nu} G_{\mu\nu}
+ \frac{g^2_1}{32\pi^2}\frac{k s_{\rm hid}}{f_{\rm hid}} 
B^{\mu\nu} B_{\mu\nu},
\eeq
the saxion can be similarly produced by gluon fusion, and it decays into photons, explaining the diphoton excess
for $f_{\rm hid} = {\cal O}(100)$\,GeV to $1$\,TeV or so. The decay constant is determined by the vacuum expectation
values of the saxions, and the similar size of the suggested decay constant and saxion mass imply that
the saxions are stabilized by tree-level potentials with couplings of order unity. The saxions generically decay
with a sizable branching fraction into axions, which may explain a broad decay width inferred by the ATLAS data.
The produced axions are considered to decay into gluons and photons, but if they have sizable couplings to
hidden photons, they may contribute to the invisible decay of the saxion.
In this case, however, one has to enhance the partial decay rate into photons. This will require a stabilization
of the saxion with couplings larger than unity.

Our scenario contains many axions and saxions with couplings suppressed by the decay constants
much lower than the conventional axion window. Some of them can be within the reach of the collider experiments
such as LHC and ILC, and in particular, we expect that similar excesses in the photon resonance search 
will appear at different energy scales.
Also, if one of the hidden axions  responsible for the diphoton excess at $750$~GeV
is actually a hidden meson that arises from a hidden sector with strong gauge interactions, additional hidden
hadrons may appear.
 On the other hand, if one of the saxions is responsible for the diphoton excess, the hidden axions
may be relatively light, 
such (relatively) light axions with decay constants of order the weak scale can be searched for
by various experiments. (See e.g.~Refs.~\cite{Jaeckel:2012yz, Fukuda:2015ana} for the mass-dependent limit
on the axion-like particle with decay constant larger than the weak scale.)
We shall comment on the cosmological constraint below, and more detailed analysis
will be presented elsewhere.

Now let us discuss cosmological implications of our scenario. The QCD axion is known to be a plausible
candidate for dark matter. If the QCD axion exists during inflation, however, it acquires quantum fluctuations
of order the Hubble parameter, leading to sizable isocurvature perturbations. The recent Planck observation~\cite{Ade:2015lrj}
placed a stringent limit on the admixture of isocurvature perturbations, setting an upper bound on the
inflation scale, $H_{\rm inf} \lesssim 10^{7}$~GeV, for $f_{\rm QCD} = {\cal O}(10^{11})$~GeV. On the other hand,
if the QCD axion is a combination of composite axions, it does not exist during inflation, and so
no isocurvature perturbations are generated. The dominant production of the QCD axion will then be from
collapse of the string-wall network, and the right amount of dark matter is produced
for $f_{\rm QCD} = {\cal O}(10^{10})$~GeV~\cite{Kawasaki:2014sqa}.

If the U(1)$_{\rm PQ}$ symmetry becomes spontaneously broken after inflation,
domain walls are formed soon after the quark-hadron phase transition. Unless the domain wall number is equal to unity, 
domain walls are long-lived and one will confront the domain wall problem. 
One of the solutions to avoid the cosmological disaster is to keep the PQ scalar fields displaced from the origin during
and after inflation by certain couplings with the inflaton.

We also note that, in our model of the QCD axion from the aligned axions,  the alignment does not necessarily 
take place for the saxions. Therefore, the saxions generically have unsuppressed couplings around $1/f$ to other particles including axions, 
and thus they are short-lived in contrast to the conventional scenario where the saxion is long-lived and tends to dominate 
the energy density of the Universe.

There are several cosmological and astrophysical constraints on the axions. One of the most stringent bounds comes from the supernovae (SN1987A).
Axions can be efficiently produced in the core of the supernovae having the extremely high temperature $T \sim 30~{\rm MeV}$ and high 
density ($\rho \sim 3 \times 10^{14}~{\rm g~cm^{-3}}$) environment.
The most efficient process for axion production is the nucleon-nucleon bremsstrahlung, $N+N \to N + N +( {\rm axion})$.
In order to be consistent with the energy-loss rate in the supernovae, axion must be weakly coupled with the nucleon, which is roughly $f \gtrsim 10^9~{\rm GeV}$, 
or the axion mass must be much larger than the supernovae core temperature.
For the decay constant of order the weak scale, the axion mass should be heavier than about  $1$~GeV~\cite{Giannotti:2010ty}. 

In this letter, we have pointed out a possibility that the QCD axion with a decay constant in the
intermediate scale (or higher) arises from multiple $(\sim 10)$ axions with decay constants $f_i$ much
lower than the conventional axion window. The decay constants $f_i$ can be as low as ${\cal O}(100)$~GeV
or TeV scale. Some of the axions or saxions may be composite particles made of hidden quarks/gauge fields where the corresponding
hidden gauge interactions are strongly coupled. In this case, those composite axions (or saxions) are actually
hidden mesons or glueballs, and their masses can be naturally comparable to the decay constant. 
Interestingly, one of such hidden composite particles can account for the recently found $\gamma \gamma$ resonance
at about $750$~GeV. If this is the case, our scenario predicts that many signals due to (composite) axions as well as other hidden hadrons 
will show up at the TeV scale in the rest of the LHC Run2.

%
\section*{Acknowledgments}
This work is supported by MEXT KAKENHI Grant Numbers 15H05889 (F.T.) and 23104008 (F.T.),
JSPS KAKENHI Grant Numbers 24740135 (F.T.),  26247042(F.T. and T.H.), and 26287039 (F.T.),
World Premier International Research Center Initiative (WPI Initiative), MEXT, Japan (F.T.),
and MEXT-Supported Program for the Strategic Research Foundation at Private Universities,
"Topological Science", Grant Number S1511006 (T.H.),
the Max-Planck-Gesellschaft, the Korea Ministry of Education, Science and Technology, 
Gyeongsangbuk-Do and Pohang City for the support of the Independent Junior Research Group at the 
Asia Pacific Center for Theoretical Physics (N.K.).
This work is also supported by the National Research Foundation of Korea (NRF) grant funded by the Korea government (MSIP) 
(NRF-2015R1D1A3A01019746) (K.S.J). 
%

\appendix

\section{Aligned QCD axion}
\subsection{A possible UV completion}
 
Here we give a possible UV completion of the aligned QCD axion
based on Refs.~\cite{Choi:2014rja,Harigaya:2014rga,Choi:2015fiu,Kaplan:2015fuy}.
We consider $N$ complex scalars $\Phi_i$ with $i = 1,2, \cdots, N$ with the following
potential,
\begin{align}
V(\{\Phi_i\}) =& \sum_{i=1}^{N} \left(
-m^2 \Phi_i^\dag \Phi_i + \frac{\lambda}{4} |\Phi_i^\dag \Phi_i|^2
\right) +  \sum_{i=1}^{N-1} \left(
\epsilon \Phi_i^\dag \Phi_{i+1}^3 + {\rm h.c.}
\right)\nonumber \\
&+\, y_q \Phi_N \sum_{\alpha = 1}^{n_q}  \bar{Q}_\alpha  Q_\alpha
+ y_\ell \Phi_N \sum_{\alpha = 1}^{n_\ell} \bar{L}_\alpha L_\alpha ,
\end{align}
where we assume that all the $\Phi_i$ develop vacuum expectation values $f_i \sim f$
of the similar size, and we introduce $n_q$ PQ quarks $Q_\alpha$ and 
$n_\ell$ PQ leptons $L_\alpha$. See the Table~\ref{tab:charge} for the 
the charge assignments of the PQ fermions. The above form of the potential is ensured by assigning
U(1)$_{\rm PQ}$ charges of ${\Phi_i}$ as $q_i = 3^{N-i}$.
Integrating out these PQ fermions leads to the effective Lagrangian (\ref{action}) with 
$n_i=3$, $k_s = n_q$ and $k = 3 a^2 n_q + b^2 n_\ell$.
Note that the diphoton excess can be explained by the hidden axion 
with $f$ around $k\times 100$~GeV while PQ fermions have masses around $f$,
implying that large $k$ is desirable. 
For instance, $f$ around or above TeV requires $k$ larger than 10, which can be obtained by PQ fermions
with an appropriate PQ and hypercharge assignment.   
The required value of $k$ is also realized without PQ leptons if the hypercharge of PQ quarks 
is sufficiently large, $a \gtrsim \sqrt{3/n_q}$.
The domain-wall number of the QCD axion is given by $n_q$, and so one needs $n_q = 1$
in order to avoid the domain-wall problem if the PQ symmetry breaking occurs after inflation.

\begin{table}[tb]
\begin{center} {\tabcolsep = 2mm
	\begin{tabular}{c|ccccc} \hline
		\rule[0mm]{0mm}{4.0mm} & $\Phi_i$ & $Q_\alpha$ & $\bar{Q}_\alpha$ & $L_\alpha$ & $\bar{L}_\alpha$  \\ \hline
		SU(3) & ${\bf 1}$ & ${\bf 3}$ & ${\bf \bar{3}}$  & {\bf 1} & {\bf 1} \\
		U(1)$_Y$ & 0 & $a$ & $-a$  & $b$ & $-b$  \\ 
		U(1)$_{\rm {PQ}}$ & $3^{N-i}$ & $-1$ & 0  & $-1$ & 0 
		\\ \hline
	\end{tabular} }
\end{center}
\caption{Charge assignment of the PQ scalars and fermions}
\label{tab:charge}
\end{table}

\subsection{High quality of the PQ symmetry}
 
It is known that the quality of the PQ symmetry must be extremely high in order for the PQ
mechanism to successfully solve the strong CP problem~\cite{Carpenter:2009zs}.
Considering that there are no exact continuous global symmetries in quatum
gravity, the PQ symmetry is considered to be explicitly broken by Planck-suppressed
operators, and therefore,
such a high quality of the PQ symmetry is a puzzle. For instance, a dimension five Planck-suppressed
operator induces an extra QCD axion mass, 
\beq
\Delta m_{\rm QCD} \sim 10^6 \,{\rm GeV} \lrfp{f_{\rm QCD}}{10^{10}{\rm \,GeV}}{3/2},
\eeq
which is about $10^{23}$ times larger than required by the successful PQ mechanism. 

In our scenario, the above puzzle can be naturally explained by the fact that all the scalars have vacuum expectation values
much smaller than the conventional axion window, and any Planck-suppressed PQ-breaking 
operators are highly suppressed.  The axion decay constant in the intermediate scale or higher
is just a mirage due to the alignment mechanism.
It is easy to see that dimension five Planck-suppressed operators are still harmful and spoil the PQ mechanism 
unless highly suppressed.
To forbid them, we impose  extra $Z_2$ parity under which $\Phi_i$ goes to $-\Phi_i$.
The $Z_2$ parity is nothing but a $Z_2$ subgroup of the U$(1)_{\rm PQ}$ symmetry.
Then one of the most dangerous Planck-suppressed operators is 
\beq
\label{dim-6}
\frac{\kappa}{6!}\frac{\Phi^6_1}{M^2_p} + {\rm h.c.}, 
\eeq
because the QCD axion comes mostly from $\arg{\Phi_1}$. 
Here $M_p \simeq 2.4 \times 10^{18}$\,GeV is the reduced Planck mass.
The above operator provides extra mass to the QCD axion
\beq
\Delta m_{\rm QCD} \sim  
0.1
\sqrt{{\rm Re}\kappa}  
\left(\frac{f_{\rm QCD}}{10^{10}{\rm GeV}} \right)
\left(\frac{f_1}{1{\rm TeV}} \right)^2
m_{\rm QCD},
\eeq
where $m_{\rm QCD}$ is the mass from the QCD instanton effects.
Also it induces small shift of the minimum of the QCD axion potential, i.e.~non-vanishing strong CP violation 
angle:   
\beq
\bar \theta \sim
10^{-10}\,
{\rm Im}\kappa  
\left(\frac{f_{\rm QCD}}{10^{10}{\rm GeV}}\right)
\left(\frac{f_1}{1{\rm TeV}}\right)^5,
\eeq 
which should be smaller than $10^{-10}$ not to generate too large neutron electric dipole moment. 
The contributions from the other dimension six operators  are comparable to or smaller than those 
from (\ref{dim-6}), and higher dimensional operators give negligible contributions. 
Thus the high quality of the PQ symmetry is naturally explained in our scenario.
It is interesting to note that the aligned QCD axion leads to testable CP violation.



\end{document}